\documentclass[conference]{IEEEtran}
\IEEEoverridecommandlockouts
\usepackage{cite}
\usepackage{amsmath,amssymb,amsfonts}
\usepackage{algpseudocode}
\usepackage{algorithm}
\usepackage{graphicx}
\usepackage{textcomp}
\usepackage{xcolor}
\usepackage{url}
\usepackage{cleveref}
\usepackage{braket}
\usepackage{url}
\usepackage{xspace}
\usepackage{makecell}
\usepackage{tikz}
\usepackage{subcaption}
\usetikzlibrary{chains,shapes.multipart}
\usetikzlibrary{shapes}
\usetikzlibrary{automata,positioning}
\usepackage{tikzscale}
\DeclareMathOperator{\Tr}{Tr}
\newcommand{\bigO}{\mathcal{O}}

\def\BibTeX{{\rm B\kern-.05em{\sc i\kern-.025em b}\kern-.08em
    T\kern-.1667em\lower.7ex\hbox{E}\kern-.125emX}}
\begin{document}

\title{KrMM Recurrences: HPC-Enhanced Estimators for Nonlinear Classical Shadows
}

\author{\IEEEauthorblockN{Matthias Gruber\IEEEauthorrefmark{1}\IEEEauthorrefmark{3}\thanks{{\IEEEauthorrefmark{3}These authors contributed equally to this work.} Corresponding authors: \texttt{\{matthias.gruber,sabrina.herbst\}@tuwien.ac.at}},
Sabrina Herbst\IEEEauthorrefmark{1}\IEEEauthorrefmark{3},
Ivona Brandic\IEEEauthorrefmark{1},
Vincenzo De Maio\IEEEauthorrefmark{2}\IEEEauthorrefmark{1}
}
\IEEEauthorblockA{
\begin{tabular}{cc}
\centering
\IEEEauthorrefmark{1}\textit{Research Unit of Computational Sustainability} & \IEEEauthorrefmark{2}\textit{School of Computing and Mathematical Sciences}\\
TU Wien, Vienna, Austria & University of Leicester, Leicester, United Kingdom\\
\end{tabular}
}
}

\maketitle

\newcommand{\trace}[1]{\operatorname{Tr}(#1)\xspace}

\begin{abstract}
While quantum technologies are rapidly evolving, near-term applications heavily rely on hybrid quantum-classical paradigms.
Among these, the classical shadows framework has gained a lot of traction, not only for quantum information science, but also for the development and verification of quantum hardware.
It allows to efficiently classically estimate many properties of quantum states from randomized measurements alone, a task that is otherwise intractable due to the destructive nature of measurements in quantum systems.
There are, however, many properties of interest that are, due to their very problem formulation, not efficiently estimable.
In this work, we identify a computational kernel occurring in estimators of polynomial functions, a class of problems that are in general computationally difficult to solve.
We show that this structure allows a row-wise partition that, together with a custom SIMD vectorization, enables massive parallelization and can reduce the computational cost significantly.
We show applicability and scalability of the approach on different tasks, and our results showcase that the approach allows high sequential throughput and close-to-ideal parallel scaling. The work thus provides a new way of identifying and exploiting structure in classical shadow estimators, resulting in practical advantages on inherently difficult problems. 
\end{abstract}


\section{Introduction}

Hybrid quantum-classical paradigms have emerged as an important means to harness and improve the capabilities of current quantum hardware.
An important application is the classical shadows formalism~\cite{Huang_2020_predicting-many-properties}, which allows deriving properties of quantum states based on randomized measurements alone, otherwise difficult due to the very nature of destructive measurements.
The protocol enables efficient unbiased estimation of many linear properties, such as expectation values of local observables, fidelity estimation of any state against a pure state, as well as entanglement verification for pure states.

Nonlinear properties, such as the purity, can be estimated, however, the resulting estimator may become computationally expensive. 
In particular, polynomial quantities are estimated using the U-statistic formalism~\cite{Huang_2020_predicting-many-properties}, i.e., we can estimate $\trace{\rho^k}$ by averaging over all size-$k$ subsets of the $N$ measurements.
This results in a computational complexity of $\bigO(N^k)$, rendering the evaluation of higher-order moments increasingly expensive. Further, many interesting properties, such as the purity~\cite{Elben_2020_mixed-state-entanglement}, have an exponential sample complexity $N$ in the number of qubits $n$, making the overall estimation procedure inefficient.
Nonetheless, such estimators are of practical interest, e.g., for quantifying entanglement~\cite{Elben_2020_mixed-state-entanglement}, thus motivating research for increased scalability.

Storing the (estimated) density matrix of a quantum state $\rho$ requires exponential-in-$n$ memory, which can be avoided by utilizing the Kronecker structure imposed by classical shadows. While used extensively for linear properties, for higher-order terms the sheer amount of operations required becomes a major practical bottleneck~\cite{marso2026onlineapproachentanglementverification}.
 
To overcome the combinatorial explosion, batch shadows~\cite{Rath_2023_batch-shadows} were proposed, where the measurements are placed into distinct batches, and estimators are built by averaging over the snapshots of a batch.
This yields an unbiased estimator with improved scalability, at the cost of storing the $N_B$ batch shadows as $2^n \times 2^n$ density matrices. 
Compared to the vanilla shadows, not all combinations are considered anymore, thus more measurements may be required for the same accuracy.

In this work, we identify a recurring computational kernel in polynomial estimators and formulate it as a chain of Kronecker-matrix-matrix (KrMM) recurrences. These recurrences require many small 
matrix products, allowing independent row-wise computations, with dependencies only between successive steps. 
This structure can be exploited for more efficient computation, e.g., through the recent {FastKron} algorithm~\cite{Jangda_2024_fastkron}, 
and a custom SIMD vectorization scheme.
\begin{figure*}[!ht]
    \centering
    \includegraphics[width=\linewidth]{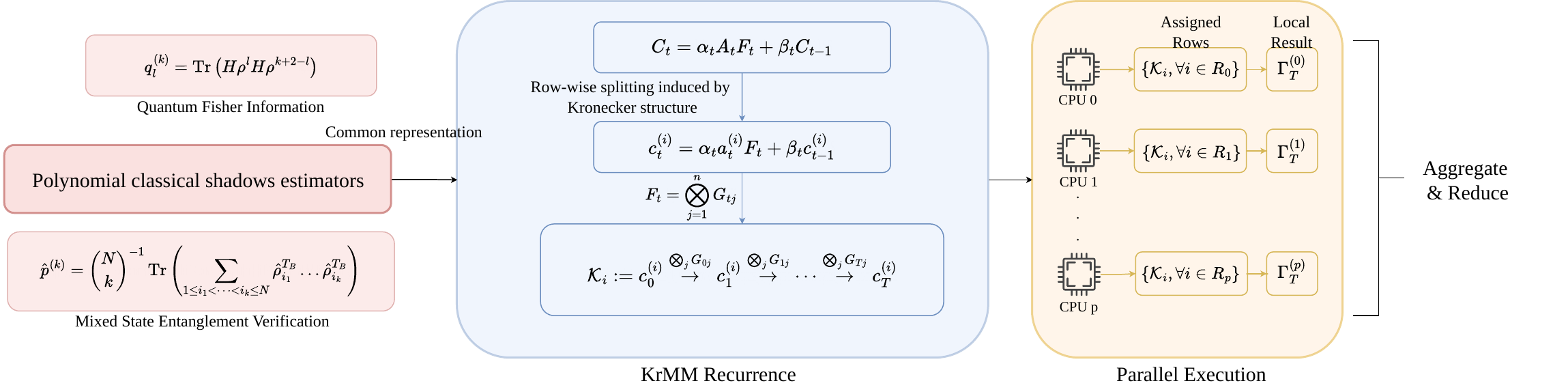}
    \caption{Our contribution. Polynomial classical shadows estimators, including online and batched~\cite{Rath_2023_batch-shadows} estimators for mixed-state entanglement verification and the Quantum Fisher Information, (see Sec.~\ref{sec:applications}), share a common representation, \textit{the KrMM recurrence}, which induces a row-wise splitting by the Kronecker structure of $F_t$. This allows defining a computational kernel $\mathcal{K}_i$, that iteratively computes the row estimators \textit{independently} of each other, as each computation only depends on the previous timestep, making the computation highly parallelizable. Each worker locally aggregates the results, and communication is only required for the final aggregation and reduction. This results in high-throughput and near-ideal parallel scaling.}
    \label{fig:krmm-overview}
\end{figure*}
We implement our approach on estimators for mixed-state entanglement verification~\cite{Elben_2020_mixed-state-entanglement} and the Quantum Fisher Information estimation~\cite{qfi-random-meas, qfi-shadows-batch, Wang_2026_KST-superiority}, as well as batch shadows.

\emph{In summary, our contributions are as follows:} (1) we identify a computational kernel underlying classical shadows estimators of polynomial functions; (2) we show that this formulation decomposes into independent row-wise calculations; and (3) we leverage this structure to massively parallelize the calculations, achieving near-ideal parallel scaling, and high throughput. Using state-of-the-art methods of parallel computing, we thus obtain a massive practical speed-up for a problem of increasing interest, that has not been parallelized on this scale so-far. Fig.~\ref{fig:krmm-overview} illustrates our approach.

The paper is structured as follows.
Sec.~\ref{sec:background} provides background information and Sec.~\ref{sec:related-work} discusses related work.
Sec.~\ref{sec:methodology} introduces the KrMM recurrence and execution model.
Sec.~\ref{sec:applications} discusses our applications and Sec.~\ref{sec:experimental-setup} the experimental setup.
Finally, we present our results in Sec.~\ref{sec:results}, discuss limitations in Sec.~\ref{sec:limitations} and conclude the paper in Sec.~\ref{sec:concl-future-work}.

\section{Background}\label{sec:background}
\subsection{Classical Descriptions of Quantum States}
Near-term quantum computing is often interested in properties of quantum states.
However, as quantum measurements are probabilistic and destructive, determining properties of arbitrary quantum states is non-trivial.
Therefore, much research has been conducted on the classical reconstruction of quantum states, i.e., quantum state tomography~\cite{Banaszek_2013, Blume-Kohout_2010, gross}.
In the most general case, it requires, however, exponential-in-$n$ many copies of the state, accompanied by exponential memory complexity and inefficient data processing.

Thus, a related line of research has considered whether, rather than reconstructing the full density matrix, it is sufficient to predict only certain properties of interest.
This problem, called shadow tomography~\cite{Aaronson_2018_shadow-tomography}, aims to predict expectation values of a fixed set of observables for a quantum state.
The circuits and measurements required by the protocol are, however, challenging to implement on near-term hardware, which motivated the development of classical shadows.

\newcommand{\setU}{\ensuremath{\mathcal U}\xspace}
\subsection{Classical Shadows}
\begin{figure*}[!ht]
    \centering
    \includegraphics[width=.9\linewidth]{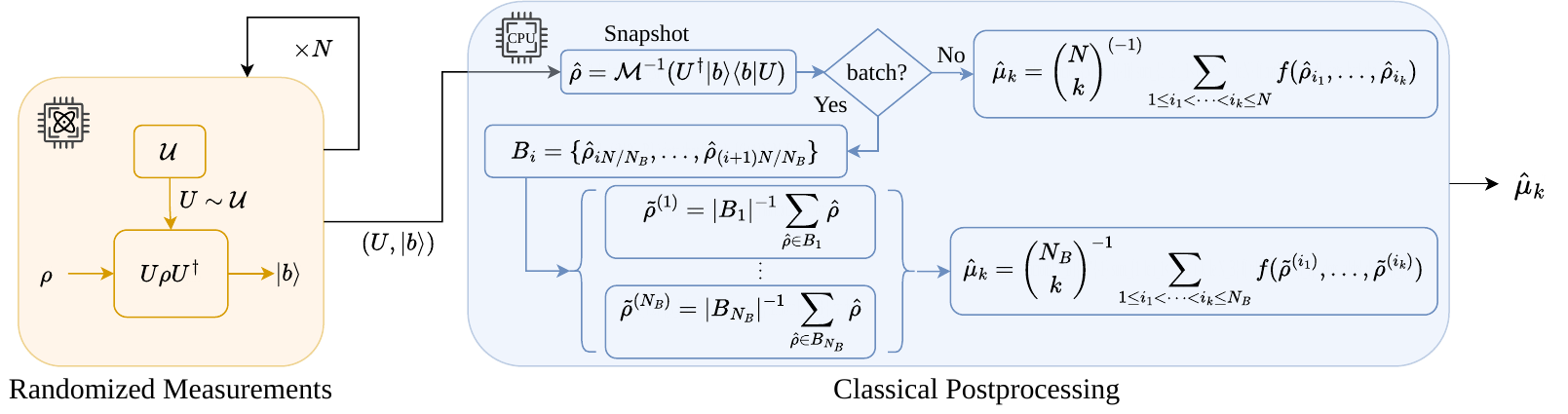}
    \caption{Overview of Plain and Batch Classical Shadows. On the left (orange), quantum measurements are collected by randomly sampling a unitary $U$ from the ensemble \setU. The process is repeated $N$ times, and the measurements $\{(U_i,\ket{b}_i)\}$ are forwarded to the classical computer for post-processing. For the plain shadows (top), the estimator for the $k$-th order polynomial is then built by iterating over all size-$k$ subsets of the snapshots, scaling as $\bigO(N^k)$. For batch shadows (bottom), the snapshots are split into $N_B \ge k$ distinct batches. Batch shadows are built by averaging over all snapshots of the batch, and the $k$-th order polynomials are estimated by iterating over all size-$k$ subset of batches. This reduces the number of combinations, at the expense of losing accuracy by not considering all cross products and storing the batch shadow density matrices in memory.
    }
    \label{fig:shadows-overview}
\end{figure*}

The classical shadows formalism~\cite{Huang_2020_predicting-many-properties} considers an unknown fixed state $\rho$, and an ensemble of unitaries \setU.
The ensemble is commonly chosen to be tensor products of single-qubit Clifford unitaries, as this makes the protocol suitable for near-term hardware.
Next, randomized measurements are taken, i.e., the state is rotated according to a random $U\sim\setU$, and measured in the computational basis.
The unitary $U$ and measurement result $\ket{b}\in\{0,1\}^n$ are then stored classically.

The average mapping of $\rho$ to the classical description can be seen as a channel, which, when inverted, gives $\rho$ as
\begin{align}
    \mathbb{E}_{b, U}[U^\dagger\ket{b}\bra{b}U] = \mathcal M(\rho) \Rightarrow \rho = \mathbb{E}[\mathcal M^{-1}(U^\dagger\ket{b}\bra{b}U)].
\end{align}
While $\mathcal M^{-1}$ is not a valid quantum channel anymore, it is efficiently computable classically for \setU, allowing, for single-qubit Clifford unitaries, to obtain an unbiased estimator of $\rho$ 
\begin{equation}
    \label{eq:snapshot-definition}
    \hat{\rho} = \bigotimes_{i=1}^{n} \left( 3 U_{i}^\dagger | b_{i} \rangle \langle b_{i} | U_{i} - \mathbb{I} \right).
\end{equation}

A classical shadow is then built as a collection of $N$ independent snapshots. We provide an overview of the formalism in Fig.~\ref{fig:shadows-overview}.
The sample complexity $N$ is generally problem dependent but can often be shown to be efficient and independent of the system size\footnote{There are many subtleties involved here, with dependencies on \setU and the observables of interest. We refer the interested reader to the original source~\cite{Huang_2020_predicting-many-properties}.}.
This allows properties to be predicted via a median-of-means algorithm, which is often possible without reconstructing the whole density matrix.

The approach naturally extends to polynomials via U-statistics formalism, where we average over products of distinct snapshots.
For quadratic terms, we would thus consider 
\begin{equation}
    \hat{o} = \frac{1}{N(N-1)} \sum_{i\ne j}\trace{O\hat{\rho}_i\otimes \hat{\rho}_j}.
\end{equation}

To estimate polynomials of order $k$, the computational complexity grows as $\mathcal O(N^k)$.
Thus, even for moderate $k$, the scaling is unfavorable, which led to batch shadows.

\subsection{Batch Shadows}
Batch shadows were originally proposed in the context of estimating operator entanglement~\cite{Rath_2023_batch-shadows}, aiming to reduce the combinatorial explosion in the estimator by dividing the classical snapshots into $N_B$ distinct batches, where $N_B \ge k$.
Denoting by $B_i$ the set of snapshots in the $i$-th batch, we can then define its batch shadow as
\begin{equation}
    \label{eq:batch-def}
    \bar{\rho}_i = |B_i|^{(-1)}\sum_{\hat{\rho} \in B_i}\hat{\rho}.
\end{equation}

We can then obtain an unbiased estimator for the observable $O^{(k)}$ by averaging over all $k$-tuples of batches as
\begin{equation}
    \hat{o} = \binom{N_B}{k}^{-1}\sum_{1 \le b_1< \dots < b_k\le N_B}\operatorname{Tr}\big(O^{(k)}\bigotimes_{i=1}^{k}\bar{\rho}_{b_i}\big).
\end{equation}

While this formulation reduces the combinatorial terms, it requires keeping the $N_B$ exponentially-sized matrices in memory. Further, as not all snapshot-level cross-products are considered anymore, more samples may be required.
See~\cite{Rath_2023_batch-shadows} for the derivation and discussion on the resulting variance, and~\cite{marso2026onlineapproachentanglementverification} for empirical results on sample complexity.

\section{Related Work}\label{sec:related-work}
The computational complexity of shadows applications is primarily governed by the choice of ensemble \setU and the observables of interest.
Many properties remain difficult to estimate, despite being of practical relevance, which has triggered research on pushing the scalability of these estimators.
In particular, recently, a formulation of the estimators as online algorithms was proposed~\cite{marso2026onlineapproachentanglementverification}, where
the estimators are updated on-the-fly instead of collecting all measurements first. 
 This leads to a truly hybrid system, where both classical and quantum resources are used simultaneously, while also allowing to monitor convergence to employ early stopping.
These practical improvements are particularly important for challenging problems such as operator and mixed-state entanglement~\cite{Rath_2023_batch-shadows, Elben_2020_mixed-state-entanglement}, the Quantum Fisher Information~\cite{qfi-random-meas,qfi-shadows-batch, Wang_2026_KST-superiority}, or expectation values of multicopy observables~\cite{mco-batch}.

In this work, we exploit the Kronecker structure of the problem. A matrix represented as $n$ $2 \times 2$ Kronecker factors can be compactly stored using only $\mathcal{O}(\log M)$ memory and reduces the cost of dense matrix multiplication from $\mathcal{O}(M^3)$ to $\mathcal{O}(M^2 log M)$, with $M=2^n$. This structure often occurs in quantum information, and is ubiquitous in classical shadows by design. 
However, it also appears in classical Gaussian processes~\cite{gp-ml, kiss-gp}, which motivated the \texttt{FastKron} algorithm~\cite{Jangda_2024_fastkron}.
It obtains advantages over existing approaches~\cite{Marc_1981_shuffle-algo, LANGVILLE2004429} in terms of contiguous data access and memory operations on certain matrices (see Sec.~\ref{sec:meth-srm}), as well as communication overhead in multi-GPU settings. To the best of our knowledge, it has not yet found applications in quantum information.

\section{Kernel Formulation and Execution}\label{sec:methodology}

\subsection{The KrMM Recurrence}

We define the Kronecker matrix-matrix (KrMM) recurrence with respect to computations over matrices as
\begin{equation}
    \label{eq:KrMM-recurrence}     
    C_t = \alpha_t A_t F_t  + \beta_t C_{t-1} \quad t=1,\dots,T,
\end{equation}
with input data $\alpha_t, \beta_t \in \mathbb{C}$, $A_t, F_t \in \mathbb{C}^{M \times M}$, and recursion matrix $C_t \in \mathbb{C}^{M \times M}$.
In this work, we consider $n$-qubit systems, thus $M=2^n$.
$F_t$ has a Kronecker structure with complex-valued Kronecker factors $G_{tj} \in \mathbb{C}^{2 \times 2}$, satisfying
\begin{equation}
F_t = \bigotimes_{j=1}^n G_{tj}.
\label{eq:kronecker-structure}
\end{equation}

Efficiently parallelizing the KrMM recurrence beyond individual matrix products is generally nontrivial, since each iteration depends on all previous ones through $C_{t-1}$ and possibly $A_t$. This is especially problematic if the matrix dimension $M$ is relatively small but the number of iterations $T$ is large, which is often the case in nonlinear classical shadow applications~\cite{Neven_2021_symmetry-resolved-entanglement, Zhou_2024_hybrid-nonlinear-functions, Zhang_2026_KST_experimental}. The Kronecker structure, however, allows a row-wise decomposition.
Denoting the $i$-th row of a matrix $B$ as $b^{(i)}$, we obtain its result through the Kronecker matrix-vector (KrMV) recurrences
\begin{equation}
    \label{eq:KrMV-recurrence}     
    c_t^{(i)} = \alpha_t a_t^{(i)} F_t  + \beta_t c_{t-1}^{(i)}.
\end{equation}
This execution model allows for an $M$-way parallelism with row kernel $\mathcal{K}_i := c_0^{(i)} \rightarrow c_1^{(i)} \rightarrow \ldots \rightarrow c_T^{(i)}$, where each iteration depends only on the shared input data $\alpha_t, \beta_t, F_t$, local input data $a_t^{(i)}$, and the row-local state $c_{t-1}^{(i)}$. Therefore, $\mathcal{K}_i \perp \mathcal{K}_j$ for $i \neq j$, allowing to map large independent computations to each worker. Note that each row kernel needs to load the shared input data, thus, practical scalability implies being able to efficiently store $F_t$ as $\mathcal{O}(\log M)$ Kronecker factors.

\subsection{Complexity and Memory Analysis}
Dense evaluation of Eq.~\eqref{eq:KrMM-recurrence} requires materializing the dense matrices and performing matrix-matrix multiplications, with a per-iteration cost of $\bigO(M^3)$.
The dominant operation in the KrMV recurrence is $a_t^{(i)}F_t$, which can be computed in $\bigO(M \log M)$ by exploiting $F_t$'s Kronecker structure. Each of the row kernels performs $T$ KrMV operations, for a total of $M\cdot T$ KrMV products and a complexity of $\mathcal{O} (T M^2 \log M)$. 

Each worker needs $\mathcal{O}(M)$ memory for executing the row kernel $\mathcal{K}_i$. If we are interested in $C_t$, we still need to store an $\mathcal{O}(M)$ output of $M$ row kernels, resulting in memory cost of $\mathcal{O}(M^2)$. 
However, often we are interested in a reduced version of $C_T$, e.g., its trace, requiring to only keep the $i$-th entry of row $i$, thus a per-row memory cost of $\mathcal{O}(1)$.
Thus, we avoid storing the dense matrix and instead only need the working memory for each row kernel, arriving at a reduced $\mathcal{O} (M)$ total memory for sequential execution, while a fully parallel setting with $p=M$ workers requires a dense $\mathcal{O} (M^2)$ memory cost distributed equally among all workers.

\begin{figure*}[t]
     \centering
     \includegraphics[width=0.8\linewidth]{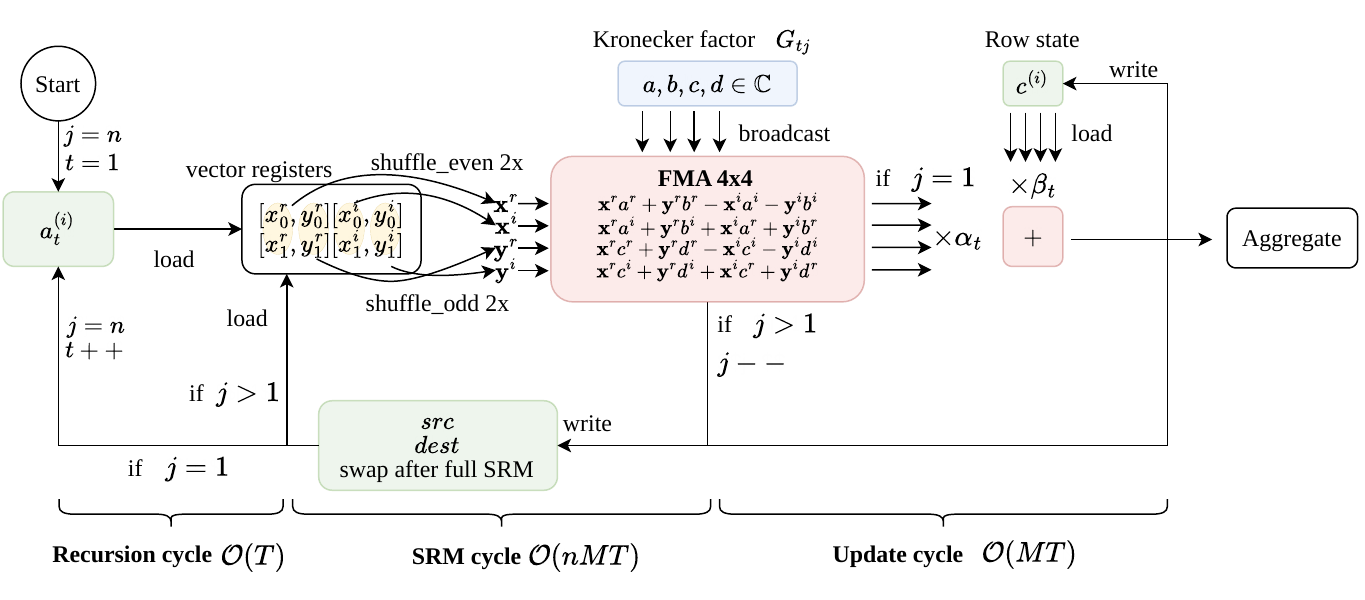}
     \caption{Fused row kernel $\mathcal{K}_i$ with custom SIMD data layout for a vector size $V_s$ of 2 doubles.
     Performing one full SRM operation (index $j$) requires fully covering one row with $M/(2V_s)$ repetitions of the SRM and update cycle. \textbf{r} indices denote real part, and \textbf{i} indices imaginary part. Green fields correspond to small data that will likely be accessed from cache. Blue fields correspond to large data that will likely be loaded from memory. In the online algorithm, each recursion cycle repeats $m$ times, with the output of the previous update cycle directly serving as the input. Aggregation in the online algorithm corresponds to storing the $i$-th entry of the row state. In the batch algorithm it corresponds to storing the entire row state in the last iteration. The combined action of shuffle + FMA preserve our data layout, enabling efficient chaining of successive SRM operations.}
     \label{fig:row-kernel}
\end{figure*}

\subsection{Sliced-Row Multiply (SRM)}\label{sec:meth-srm}
\label{sec:SRM}
Within each row kernel $\mathcal{K}_i$, the dominant operations are the KrMV products. This puts our approach close to state-vector simulation methods~\cite{Haener_2017_05petabyte, state-vector-ARM_Shi_2026}, where one is also concerned with the action of a Kronecker-structured matrix on a vector. The notable difference is that in state-vector simulation, one aims to distribute large KrMV operations over many workers, while here each worker computes many small KrMV operations in sequence, thus shifting the focus from single KrMV operations to efficiently fusing successive KrMV operations.

Therefore, we adopt the Sliced-Row Multiply (SRM) operation defined in the recent \texttt{FastKron} algorithm~\cite{Jangda_2024_fastkron}. Given a Kronecker factor $G_{tj} \in \mathbb{C}^{2\times2}$, SRM partitions the input row into slices of even and odd elements and independently applies $G_{tj}$. 
Subsequently, the complete KrMV product can be evaluated as a sequence of $n$ SRM operations, without constructing the corresponding $M\times M$ Kronecker matrix $F_t$. 

Moreover, the SRM operation requires only contiguous data access in a highly regular, predictable pattern and is therefore well suited for fusing subsequent operations. This can be seen as a specialization of the approach employed in state-vector simulation methods. Using SRM, we lose the ability to exploit structure within the Kronecker factors, a crucial aspect of state-vector simulation, but gain higher regularity and more effective fusion of successive KrMV operations. Note that since the shadow snapshots do not contain identity Kronecker factors, their structure could not be significantly exploited either way.

The resulting KrMV evaluation serves as the basic building block of every row kernel. However, SRM cannot leverage existing linear algebra routines, so we develop a custom SRM implementation with a SIMD vectorization scheme.

\subsection{Execution Model}
A conventional evaluation requires computing each iteration of Eq.~\eqref{eq:KrMV-recurrence} independently, therefore risking materializing working data to memory. Thus, we treat each row kernel $\mathcal{K}_i$
as a single fused computation, illustrated in Fig.~\ref{fig:row-kernel}.
We instantiate vector registers using only consecutive vector loads and perform as many computations on them as possible before writing them back. The state produced is used immediately in the next cycle, avoiding intermediate memory access and reducing memory traffic. 
This provides contiguous memory accesses, leading to a cache-aware implementation. 

When computing a set of coupled KrMM recurrences, the state produced by the update cycle is directly used as the input row of the next recursion cycle. 
Also, the recurrence fusion is independent of the Kronecker acceleration: while the latter reduces the cost of KrMV, the fusion reduces the cost of moving data through the recurrence. 
The row-wise decomposition naturally exposes parallelism as row kernels share no state, enabling independent distribution across the $p$ workers. Let $R_j$ be the set of rows assigned to worker $j$. Each worker $j$ evaluates all kernels $\mathcal{K}_i$, $i \in R_j$, and aggregates their results.  
Since no communication is required during the kernel execution, a single global reduction combines the local results. 

Further, as all row kernels perform the same
operations, a static row partitioning strategy naturally provides load balancing. The absence of communication during kernel execution allows near-ideal parallel scalability until bandwidth limitations become dominant. Whether bandwidth contention limits scalability hinges on the rows fitting into cache, which is usually the case for the system sizes considered in nonlinear classical shadows ($n \lesssim 12$)~\cite{Neven_2021_symmetry-resolved-entanglement, Zhou_2024_hybrid-nonlinear-functions, Zhang_2026_KST_experimental}. 

\subsection{Data Layout and SIMD Realization}
The SRM operation applies the same transformation to each row, therefore the most natural SIMD strategy is vectorizing across rows. However, our execution model assigns each row to different workers and exploits row-level parallelism, making it incompatible with cross-row vectorization. Thus, we implement a column-based SIMD strategy on a custom data layout using architecture-specific permute intrinsics.  
\par
One SRM cycle with Kronecker factor
\begin{equation}
G := \begin{pmatrix}
a & c \\
b & d
\end{pmatrix} \in \mathbb{C}^{2 \times 2}
\label{eq:vec0}
\end{equation}
transforms the row buffers of Fig.~\ref{fig:row-kernel} as \texttt{src} $\xrightarrow{G}$ \texttt{dest}. In the following we denote by $x_l, y_l \in \mathbb{C} , l<M/2$ even and odd entries of \texttt{src} in the natural data layout of interleaved real and imaginary parts.
Every result element of \texttt{dest} is obtained by summing up an even-odd pair $x_l, y_l$, specifically
\begin{equation}
\begin{aligned}
\texttt{dest}[l] &= x_{l} \cdot a + y_{l} \cdot b \\
\texttt{dest}[M/2 + l] &= x_{l} \cdot c + y_{l} \cdot d. \\
\end{aligned}
\label{eq:vec2}
\end{equation}
Efficient vectorization requires extracting $x_l$ and $y_l$ into separate SIMD registers using only consecutive vector loads, while also supporting complex FMA arithmetic. Furthermore, SRM operations are evaluated successively, so the output of one operation must serve as input to 
the next Kronecker factor without additional
reorganization. Thus, we use a data layout where we group $V_s$ real and $V_s$ imaginary floats together, visible in Fig.~\ref{fig:row-kernel} for a vector size of 
$V_s = 2$.
This allows extracting the data using \texttt{shuffle\_even} and \texttt{shuffle\_odd} primitives, which, given two SIMD vectors, fill a new vector with all their even or odd entries.
These operations are implemented using architecture-specific permute intrinsics, enabling vectorized complex FMA operations while preserving the layout.
 
The kernel performs a 4-way pattern of 1 load, 1 shuffle, 4 FMA, and 1 write, where the efficiency depends on the architecture support for intrinsics permuting arbitrary elements of two input vectors to one output vector. 
In practice, the CPU can efficiently hide the cost if native support is given, as in e.g. AVX-512\footnote{For AVX-512 we use the SIMD intrinsic \texttt{\_mm512\_permutex2var\_pd}.}. Relying on a composition of multiple intrinsics, e.g., for AVX2\footnote{For AVX2 we use two \texttt{\_mm256\_permute4x64\_pd} followed by one \texttt{\_mm256\_blend\_pd}.}, leads to a noticeable performance loss. The data layout is extendable to all power-of-2 vector sizes $V_S$ as long as it fits into one row at least two times, i.e. $2V_S \leq M$.

\section{Applications}\label{sec:applications}

Beyond describing the formalism for general polynomial functions, we showcase our approach for two specific estimators: 
mixed state entanglement verification and Quantum Fisher Information estimation, which represent practically relevant
problems (e.g., quantum-state verification~\cite{Horodecki-ppt} and quantum metrology~\cite{qfi}) and allow for online and batch estimators that can be expressed via the KrMM recurrence.

\subsection{Mixed State Entanglement Verification}
Entanglement of mixed states can be verified using the positive partial transpose~\cite{Horodecki-ppt, Horodecki-ppt2, ppt3} condition.
In a nutshell, for a state $\rho$ on partitions $A$ and $B$, negative eigenvalues of the partially transposed density matrix suffice to establish that $A$ and $B$ are entangled, however, this is computationally difficult as it requires access to the full density matrix.
An alternative criterion can be found via the moments of the partial transpose, which can be estimated using classical shadows~\cite{Elben_2020_mixed-state-entanglement, Neven_2021_symmetry-resolved-entanglement}, but become increasingly computationally difficult with higher-order moments.
We apply the KrMM recurrence on two estimators; the online estimator from~\cite{marso2026onlineapproachentanglementverification} and batch shadows.

\paragraph{Online Algorithm} 
The $k$-th moment at shot $t\in[1,\dots,N]$ is defined as 
\begin{equation}\label{eq:running-p}
    \hat{p}^{(k)}_t = \binom{t}{k}^{-1}\trace{C_{k}(t)},
\end{equation}
with the matrices $C_k(t)$ as the combinatorial sum of all $t$ snapshots with partial transposition on subsystem $B$
\begin{equation}\label{eq:running-A}
    C_{k}(t) = \sum_{1\le i_1 < \cdots < i_k\le t} \hat{\rho}_{i_1}^{T_B}\cdots \hat{\rho}_{i_k}^{T_B}.
\end{equation}
It has been shown that the matrices $C_k$ can be computed using a recurrence of Kronecker matrix products~\cite{marso2026onlineapproachentanglementverification}, which is a natural fit for our KrMM recurrence formalism.
We slightly 
adapt the original formulation, by integrating the normalization into the running estimate for numerical stability and altering the recursive structure for better kernel fusion, obtaining 
\begin{equation}
    \label{eq:online-recurrence}
    C_k(t) = \alpha_t^{k}C_{k-1}(t)\hat{\rho}_{t+k-1}^{T_B} + \beta_t^{k}C_k(t-1),
\end{equation}
with $\alpha_t^k = \frac{k}{t+k-1}$ and $\beta_t^k = \frac{t-1}{t+k-1}$ for $k = 1,\dots,m$, and initial conditions $C_k(0) = 0$ and $C_0(t) = \mathbb{I}$. 
The running moment estimate of all orders $k$ can then obtained in every iteration as $\hat{p}_{t+k-1}^{k} = \trace{C_k(t)}$. Within each $t$ iteration, the output of iteration $k$ directly serves as the input to $k+1$, which is exploited in the row kernel to reduce memory traffic.

Cheap access to the recursively built intermediate values allows uncertainty estimation via standard formula of the U-statistics variance~\cite{Hoeffding}, which we briefly derive in the following. In the first order of $t$ it is given as 
 \begin{equation}
     \label{eq:ustat-var}
     \mathrm{Var} (\hat{p}_t^k) = \frac{k^2 \sigma_1^2}{t}, 
 \end{equation}
which dominates all other terms scaling as $t^{-x}$ with $x\geq2$ for large enough $t$. The variance of the conditional expectation of the U-statistics kernel, given one input is fixed, is defined as
\begin{equation}
    \sigma_1^2 = \mathrm{Var} \left(\mathbb{E}\left[\Tr (\hat{\rho}_{i_1}^{T_B}\cdots \hat{\rho}_{i_k}^{T_B}) | \hat{\rho}_{i_k}^{T_B} \right]\right),
\end{equation}
which is a by-product of Eq.~\eqref{eq:online-recurrence}.
Subtracting $C_k(t-1)$ from both sides of the recurrence and taking the trace, together with the definition of the matrices $C_k(t)$, gives
\begin{equation}
    \frac{\binom{t}{k}}{\binom{t-1}{k-1}} \hat{p}_t^k - \frac{\binom{t-1}{k}}{\binom{t-1}{k-1}} \hat{p}_{t-1}^k = \mathbb{E} \left[ \Tr (\hat{\rho}_{i_1}^{T_B}\cdots \hat{\rho}_{i_k}^{T_B}) | \hat{\rho}_{i_k}^{T_B} \right].
\end{equation}
Computing many of these conditional expectation values at consecutive $t$ allows computing their sample variance, and thus $\sigma_1^2$ and $\mathrm{Var} (\hat{p}_t^k)$. This in turn can be used to check for convergence at a desired accuracy or propagate uncertainty to derived quantities. Extracting all intermediate moment estimates (and uncertainty) costs $\bigO(m N 2^n)$, leaving the total complexity of $\mathcal O(mnN4^n)$ unchanged. As a result, the online algorithm allows for a truly hybrid approach, where incoming measurements are immediately processed at $\bigO(mn4^n)$ cost, providing live insight into the quantum system.

\paragraph{Batch Algorithm} Computing a batch shadow $\Bar{\rho}_i$ corresponds to averaging over a set of snapshots $\hat{\rho}_t \in B_i$. This can be formulated as a KrMM recurrence using a fixed $A_t = \mathbb{I}$ and a running average, giving the defining recurrence as
\begin{equation}
    \label{eq:batched-recurrence}
    C(t) = \alpha_t\mathbb{I}\hat{\rho}_t^{T_B} + \beta_t C(t-1),
\end{equation}
with $\alpha_t = \frac{1}{t} $, $\beta_t = \frac{t-1}{t}$, $\hat{\rho}_t \in B_i$, and initial condition $C(0)=0$. 
For $N_B=k$ independent batches covering all measurements, this gives the final estimate as $\hat{p}_{N}^{k} = \trace{\prod_{i=1}^{k} \bar{\rho}_i}$.

The cost of forming the batch shadows thus scales as $\mathcal O(nN4^n)$, whereas their multiplication 
scales as $\mathcal O(m8^n)$.
However, as $N \ggg 2^n$, the dominant cost in practice is the formation of the batches, thus, we only build the batches once and estimate all $m$ moments up from them.
For this we initially build $m$ batches and compute $\hat{p}^m$. We then combine batches $m$ and $m-1$, and compute the estimate of $\hat{p}^{m-1}$. We repeat this until two batches are left for a total cost of $\mathcal O(m^2 8^n)$ and combine the batches evenly to retain the most combinations possible. 
While the KrMM recurrence serves to parallelize the computation of a single batch, the computation can be further parallelized by forming the batches concurrently.

With a total cost of $\mathcal O(nN4^n)$, the batched algorithm provides a benefit of $\mathcal O(m)$ over the online algorithm in an offline setting.
While batch formation is easily implemented online,
extracting moment estimates comes at a cost of $\mathcal O(m^2 8^n)$. For the typically small $n$ used in classical shadows, paying this once is negligible, but in an online setting, this would scale as $\mathcal O(m^2 N 8^n)$, which makes a comparison between the approaches difficult, as they cover different problem domains.

\subsection{Quantum Fisher Information}
The Quantum Fisher Information (QFI)~\cite{qfi, qfi2} was originally proposed in quantum metrology, however, has since found applications in quantum information~\cite{qfi3}. 
Consider a quantum state $\rho = \sum_i \lambda_i\ket{i}\bra{i}$ and a Hermitian operator $H$. 
The QFI is then defined as 
\begin{equation}
    F_Q = 2\sum_{(i,j)\\ \lambda_i + \lambda_j > 0}\frac{(\lambda_i - \lambda_j)^2}{\lambda_i + \lambda_j}|\bra{i}H\ket{j}|^2.
\end{equation}
While it cannot be directly expressed as a polynomial function, polynomial lower bounds can be estimated via classical shadows~\cite{qfi-random-meas}, however, the required number of bounds make it prohibitive in practice.

Thus, an alternative framing based on the Krylov subspace method
was proposed recently as Krylov Shadow Tomography (KST)~\cite{Zhang_2025_KST, Wang_2026_KST-superiority, Zhang_2026_KST_experimental}. Here, increasingly accurate non-polynomial lower bounds of the QFI are obtained from multiple polynomial quantities. Intriguingly, for a state $\rho$ and operator $H$, an analytical maximum bound order $N^*$ exists, for which the estimate and $F_Q$ coincide. Obtaining an unbiased estimate for $F_Q$ then reduces to estimating a set of polynomials
\begin{equation}
    \label{eq:QFI-polys}
    q_l^{(k)} = \Tr (H \rho^l H \rho^{k+2-l}),
\end{equation}
for $k=0,\dots,2N^*-1$ and $l=0,\dots,k+2$.
This can be computed with batch shadows by inserting the computed batches into Eq.~\eqref{eq:QFI-polys}, similar to previous works~\cite{Wang_2026_KST-superiority, Zhang_2026_KST_experimental}. We can thus directly apply the batch algorithm from Eq.~\eqref{eq:batched-recurrence}.

To obtain a fine-grained estimation of the QFI, 
we can also employ the online algorithm to estimate the polynomials $q_l^{(k)}$. For this, we use $N^*+1$ independent runs of order $m=2N^*+1$, each estimating $q_l^{(k)}$ for a fixed value of $l = 0, \dots, N^*$ and all $k=1,\dots,m$. The remaining values can be obtained by the cyclicity of the trace. To consider the two Hermitians $H$ present in the definition of $q_l^{(k)}$, we adapt the online recurrence from Eq.~\eqref{eq:online-recurrence}. The leftmost $H$ in Eq.~\eqref{eq:QFI-polys} enters as the initial condition $C_0(t)=H$. For the second $H$, we split the recurrence into two separate cases as
\begin{equation}
    \label{eq:online-recurrence-qfi-1}
    C_k(t) = \alpha_t^{k}C_{k-1}(t)\hat{\rho}_{t+k-1} + \beta_t^{k}C_k(t-1) \ , \ \ k \neq l+1
\end{equation}
\begin{equation}
    \label{eq:online-recurrence-qfi-2}
C_{k}(t) = \alpha_t^{k} C_{k-1}(t) H \hat{\rho}_{t+k-1} + \beta_t^{k} C_{k}(t-1) \ , \ \ k = l+1.
\end{equation}
We thus require
$H$ as (a sum of) Kronecker products, which is the case in many problems of interest. 
Here we consider $H = \frac{1}{2} \sum_{j=1}^n \sigma_Z^{(j)}$
, where $\sigma_Z^{(j)}$ is the Pauli-Z operator acting on qubit $j$. This Hamiltonian is useful for entanglement detection and is the same as in the main works on KST~\cite{Zhang_2025_KST, Wang_2026_KST-superiority, Zhang_2026_KST_experimental}.

\section{Experimental Setup}\label{sec:experimental-setup}

\paragraph{Baseline}

We compare our implementation to the state-of-the-art Julia package \texttt{RandomMeas.jl}~\cite{Elben_2026_arxiv_shadows-julia}, which supports batch shadows out of the box. It partitions all measurements as $N=N_U \cdot N_M$, averaging over $N_M$ measurement outcomes grouped under one of the $N_U$ unitaries by computing the empirical Born distribution. This changes the dominant cost factor from $N$ to $N_U$. To obtain the full randomized measurement statistics, we use all possible unique unitaries $N_U=3^n$ with $N_M=N/N_U$ outcomes each, and distribute the unitaries equally among all batches. To obtain a fair comparison, we compute all batches in parallel. We extract the moment estimates from the computed batches using the same scheme as in our batch algorithm.

\paragraph{Quantum States}
We use three different states with fixed experimental parameters. Unless mentioned otherwise, results provided for a given state refers to the parameters listed here. We obtain the classical shadows data and ground truth values by numerically simulating the data acquisition phase.

Werner states~\cite{werner} are a family of states, where the entanglement is controlled by a parameter $t$. We collect $N = 6\cdot 10^8$ measurements for $n=2,4,6,8,10$ from a state with $t=0.95$, requiring moments up to order $m=34$ to detect entanglement~\cite{marso2026onlineapproachentanglementverification}. These parameters were chosen to be experimentally feasible, while demonstrating the scale of our approach; for $n=10$, applying the online algorithm costs about $3$ Exaflops in total.  We use this state for entanglement verification, with the partial transpose on an equal split.
To show distributed scalability, we employ $2$ and $8$ nodes for $n=8$ and $10$ respectively for the online algorithm. 
We do not distribute the batch one, as it already outperforms the online algorithm on a single node and relies on the same implementation, thus, would have provided only limited additional insights.

The considered quantum state itself does not affect runtime in any way. To analyze performance metrics, we use the maximally mixed state ($\frac{1}{M}\mathbb{I}$), as measurement data can be simulated easily. We use $11$ independent samples with $N = 10^7$ measurements each for $n=2,\dots,8$ qubits with order $m=34$ to obtain a meaningful runtime estimate, that is not dominated by start-up costs.
For higher $n=9,\dots,12$, we decrease the number of measurements and use $11$ samples with $N=5 \cdot 10^4$ each. Results presented using this state correspond to the median out of all $11$ samples.

The random rank $2,4,6$ states refer to randomly generated states with fixed rank, used for QFI estimation. We use $11$ independent samples with $N=3\cdot10^6$ each for $n=2,\dots,8$ qubits. Results presented using this state correspond to the median out of all 11 samples.

\begin{table}[t]
    \centering
    \scriptsize
    \setlength{\tabcolsep}{7.5pt}
    \renewcommand{\arraystretch}{0.9}
    \caption{Hardware configurations.}
    \label{tab:hardware-software}
    \begin{tabular}{lccc}
        \hline
        & \textbf{EPYC 7713} & \textbf{EPYC 9354P} & \textbf{EPYC 9255} \\
        \hline
        Generation & Zen 3\cite{amd_sog_zen3} & Zen 4\cite{amd_sog_zen4} & Zen 5\cite{amd_sog_zen5} \\
        Cores               & 64   & 32    & 24    \\
        Threads               & 128   & 64    & 48    \\
        Sockets            & 2 & 1 & 1     \\
        Base               & 2.00 GHz & 3.25 GHz & 3.20 GHz \\
        Boost              & 3.72 GHz & 3.80 GHz & 4.32 GHz \\
        L1 (i/d)          & 32/32 KB & 32/32 KB & 32/48 KB \\
        L2                 & 512 KB & 1 MB & 1 MB \\
        Memory          & 512 GB DDR4 & 503 GB DDR5 & 755 GB DDR5 \\
        Max SIMD ISA        & AVX2      & AVX-512   & AVX-512   \\
        Vector width        & 32B      & 32B$^\ast$ & 64B \\
        FMA pipes           &  2     & 2 &  2 \\
        \hline
        Compiler           & AOCC 4.1.0 & GCC 15.3.0 & GCC 15.3.0 \\
        MPI                & OpenMPI 4.1.4 & --- & --- \\
        \hline
    \end{tabular}
    \vspace{2pt}
    {\\ \scriptsize $^\ast$Zen 4 implements AVX-512 via double-pumped 256-bit execution units.}    
\end{table}

\paragraph{Hardware and Software}
We benchmark our approach on three different AMD EPYC processors with the most recent microarchitecture generations Zen 3, 4, and 5, which we compare in Tab.~\ref{tab:hardware-software}. The Zen 3 configuration corresponds to a single node of an institutional HPC cluster. For multi-node experiments, we use nodes interconnected with a Mellanox HDR InfiniBand fabric (200 Gbit/s).
\par
\begin{figure}[t]
\centerline{\includegraphics[width=0.375\textwidth]{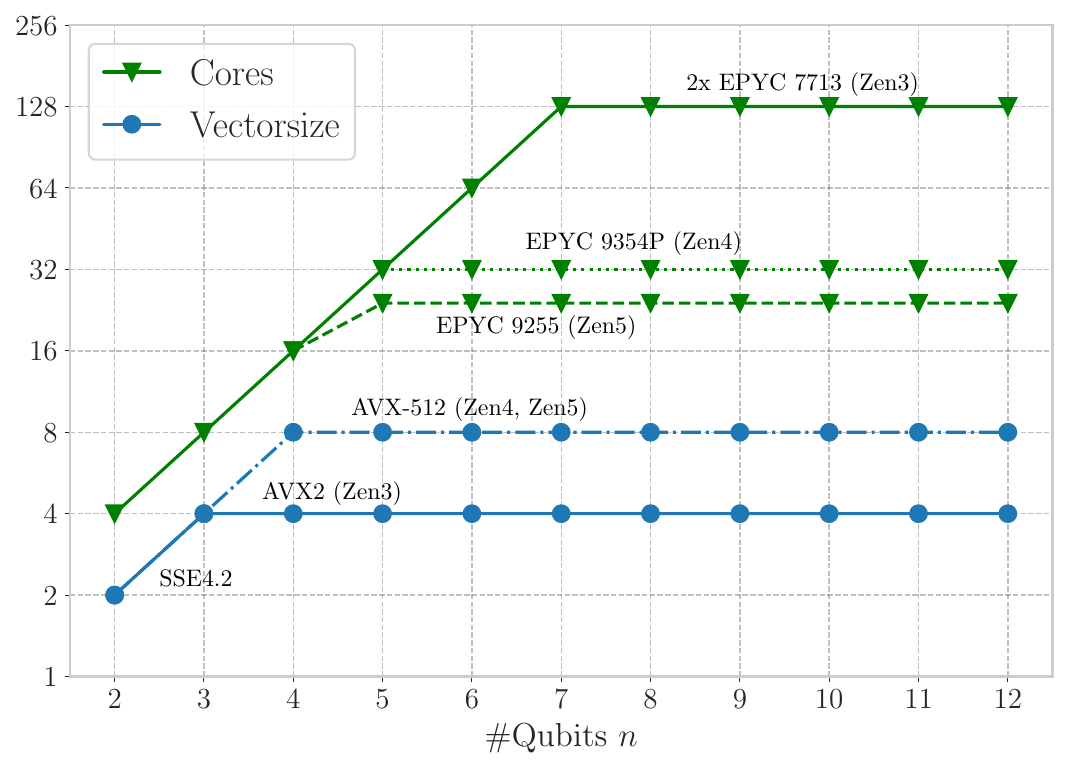}}
\caption{Used single node parallelism over increasing number of qubits. We use simultaneous multithreading (SMT) once the number of rows exceeds the number of available cores.}
\label{fig:par_specs}
\end{figure}

 
We compile using the C++17 standard with optimization flags \texttt{-O3}, \texttt{-march=native}, \texttt{-ffast-math}, \texttt{-ffp-contract=fast}, and use OpenMP~\cite{openmp} for shared memory parallelism. To implement the custom SIMD vectorization scheme, we use vector intrinsics provided by the Vector Class Library (VCL), version 2.02.01~\cite{agner_fog_vcl}. In all experiments, we use double precision floating point numbers.
The number of qubits $n$ heavily dictates the parallelism in our partitioning scheme. Unless stated otherwise, for each $n$ we use the maximum core count and vector size. Fig.~\ref{fig:par_specs} illustrates this in a single node setting.

Our source code and experimental data is accessible at~\cite{repo}.

\section{Results}\label{sec:results}
\subsection{PT-moments estimation}

\begin{figure*}[]
\centerline{\includegraphics[width=1.0\textwidth]{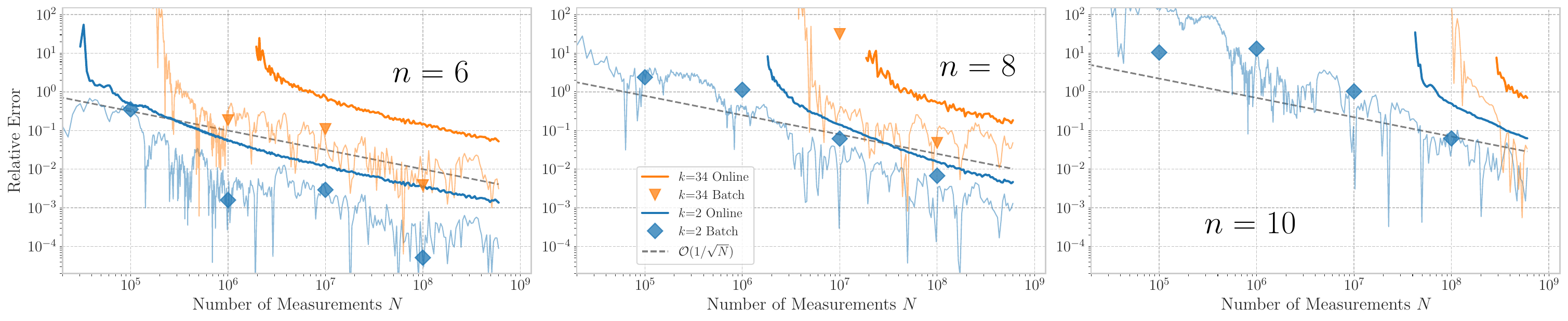}}
\caption{Relative error of PT moment estimation over number of measurements of the Werner state for moments $2$ and $34$ for 6, 8, and 10 qubits.
Lightened lines and dots depict the relative errors. 
Thick lines correspond to the probabilistic error bound of the online algorithm obtained through variance estimation.
}
\label{fig:werner}
\end{figure*}

\begin{figure}[]
\centerline{\includegraphics[width=0.37\textwidth]{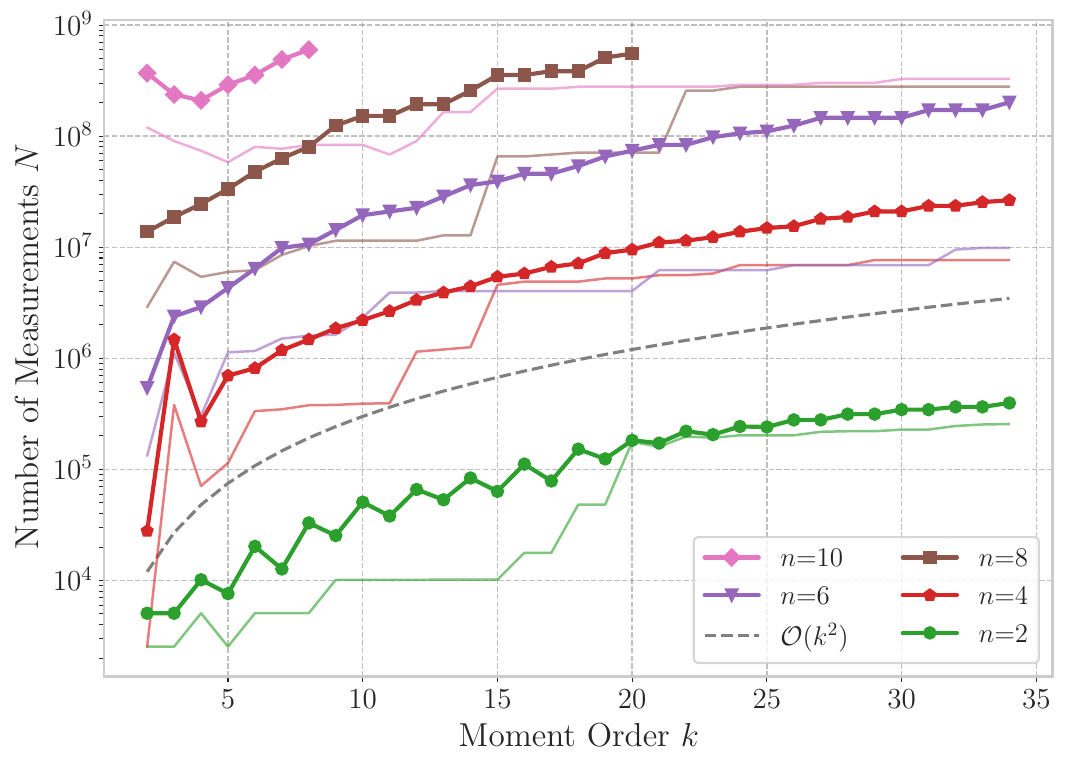}}
\caption{Number of measurements $N$ needed to reach convergence using the online variance estimation applied to PT moment estimation of the Werner state. We declare convergence once we are 95\% confident that the relative error $<10^{-1}$.
Lightened lines correspond to $N$ needed for the true relative error $<10^{-1}$ for the current, and all following data points.}
\label{fig:werner_convergence}
\end{figure}

\paragraph{Accuracy and Convergence}
To test our approach for mixed state entanglement verification, we predict PT moments of the Werner state and analyze accuracy by investigating the relative error. 
Fig.~\ref{fig:werner} shows convergence over the number of measurements for $n=6,8,10$ qubits for moment orders $2$ and $34$ for both algorithms with the KrMM recurrence. 

The online estimator allows for fine-grained convergence resolution, while the batched estimator only provides estimates at discrete evaluation points. Beyond stochastic noise, the relative error of the two algorithms are similar, which coincides with the analysis of~\cite{Rath_2023_batch-shadows}. For large enough $N$, they converge as $1/\sqrt{N}$, agreeing with the results of~\cite{Elben_2020_mixed-state-entanglement}.

In a real experimental setting, where the true moment values would be unknown, only the probabilistic error bounds (thick lines) are accessible, which only become useful
after a critical number of measurements. Before that, the confidence interval includes $0$, making the upper error bound diverge. This is not an inherent flaw of the method but simply means that it is not possible to even decide on the sign of the estimate with confidence. This accuracy estimate makes the online algorithm attractive in practice, where quantum measurements arrive one after another, and we wish to stop at a certain accuracy\footnote{Using the Jackknife procedure~\cite{Rath_2023_batch-shadows}, uncertainty estimates are also available for batch shadows, however, they are not implemented here.}. 

For a more comprehensive convergence analysis, we investigate the number of measurements needed to declare convergence with $95\%$ confidence for an accuracy of $10^{-1}$ in Fig.~\ref{fig:werner_convergence}. Since we estimate the accuracy with a probabilistic upper bound, the number of measurements to declare convergence is naturally higher than what would be needed to actually reach convergence. The measurements scale as $\mathcal{O}(k^2)$, which agrees with Eq.~\eqref{eq:ustat-var}. The exponentially many measurements $N$ needed to reach convergence make nonlinear classical shadows inherently limited to small quantum system sizes $n$, since $N$ iterations of the KrMM recurrence are required. Note, however, that this highly depends on the state and observable. 

\begin{figure}[]
\centerline{\includegraphics[width=0.37\textwidth]{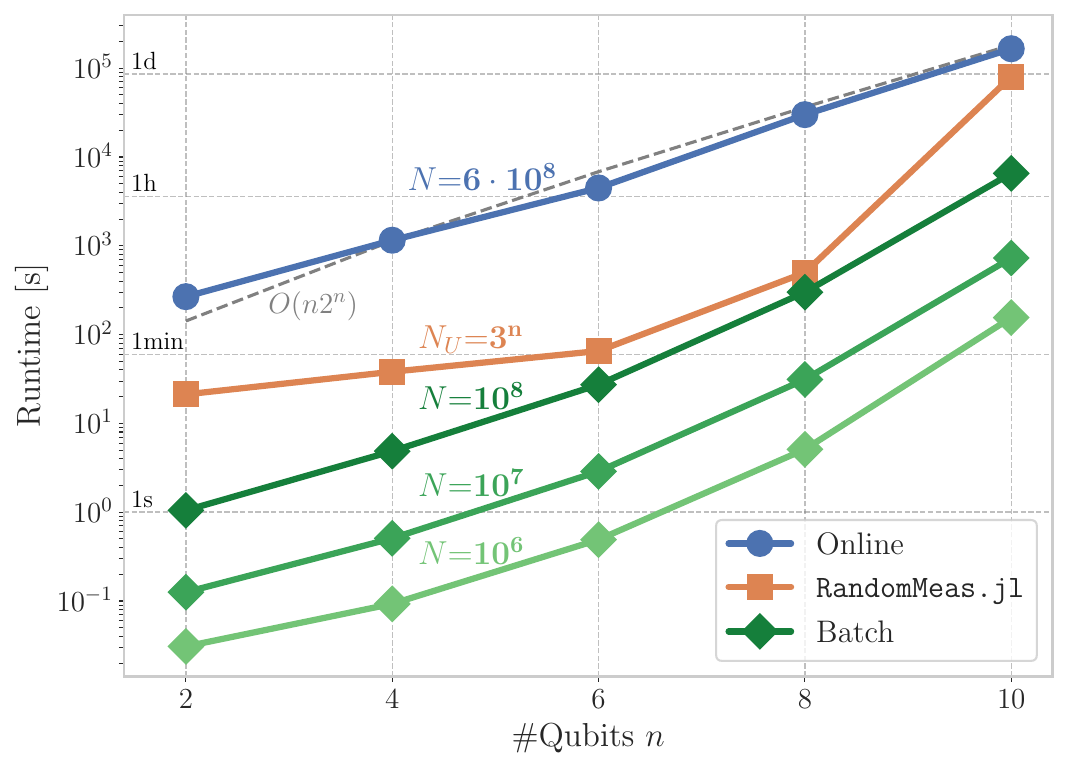}}
\caption{Runtimes of different algorithms applied to the Werner state. For the online algorithm we used $2$ nodes at $n=8$ and $8$ nodes at $n=10$. \texttt{RandomMeas.jl} and the batch algorithm were performed on a single node. For \texttt{RandomMeas.jl} we used $N_M=10^7/N_U$. Performed on the Zen 3 configuration.}
\label{fig:werner_runtimes}
\end{figure}

\begin{figure*}[t!]
    \centering
    \begin{subfigure}[t]{0.325\textwidth}
        \centering
        \includegraphics[width=1\textwidth]{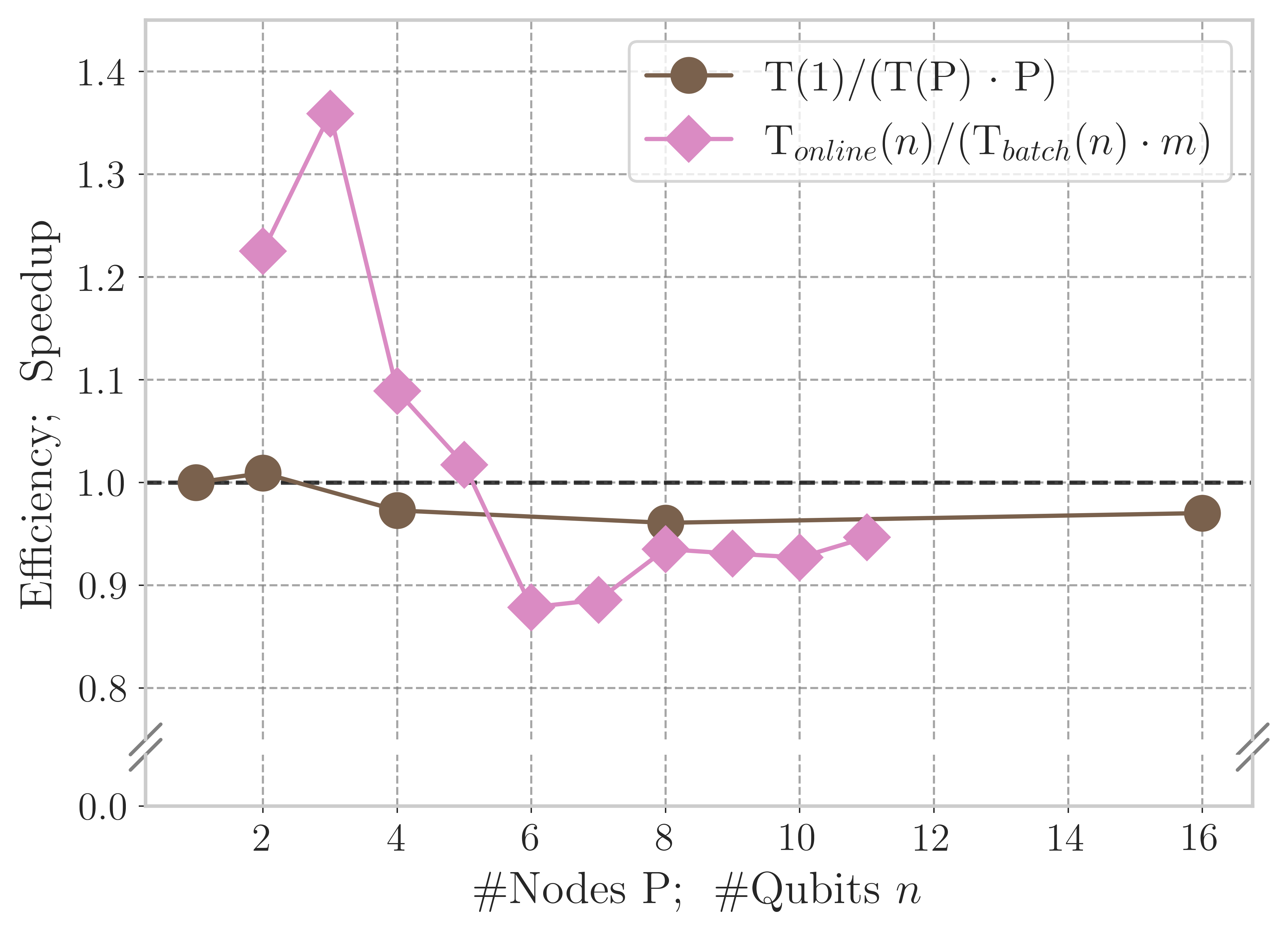}
        \label{fig:ss-eff_online-to-batch}
    \end{subfigure}
    ~ 
    \begin{subfigure}[t]{0.325\textwidth}
        \centering
        \includegraphics[width=1\textwidth]{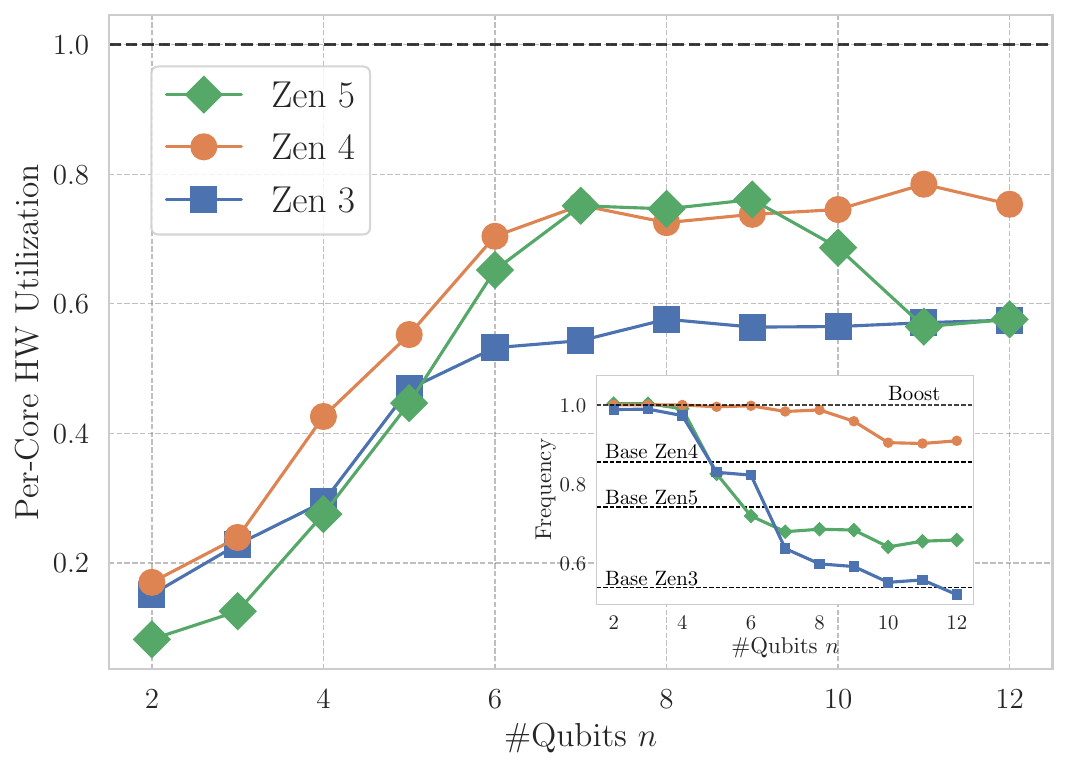}
        \label{fig:hw-util}
    \end{subfigure}
    \begin{subfigure}[t]{0.325\textwidth}
        \centering
        \includegraphics[width=1\textwidth]{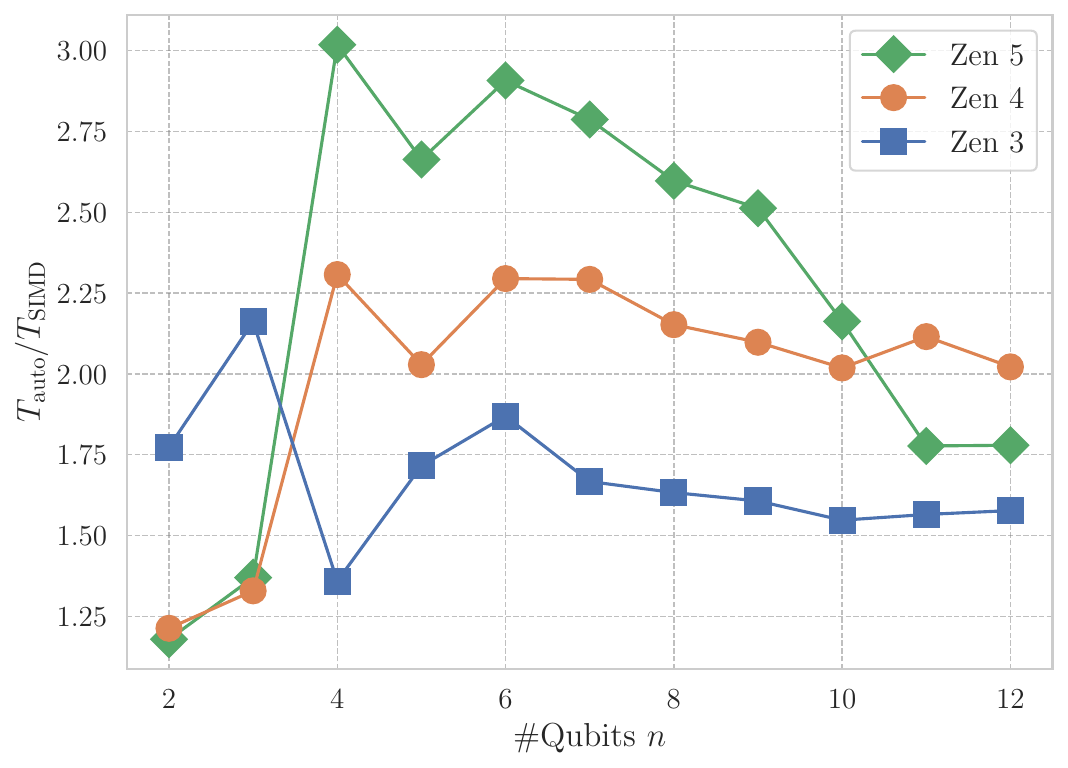}
        \label{fig:simd-su}
    \end{subfigure}
    \caption{\textbf{(left)} (brown) Strong-scaling efficiency of the distributed online algorithm for $12$ qubits using the maximally mixed state. To have a sufficiently large workload, we use $N=10^5$ measurements. (pink) Speedup of $m$ times the batch KrMM kernel to the online kernel using the maximally mixed state. Performed on the Zen 3 configuration. \textbf{(middle)} Per-core hardware utilization (FMA) of the online kernel for different microarchitectures (for number of cores, cf. Fig.~\ref{fig:par_specs}) using the maximally mixed state (single node). The CPU frequency was measured at runtime to faithfully estimate peak performance at each core count. At $n=12$ (all cores) we achieve throughputs of $2.28$, $1.33$, $1.26$ TFLOPs for Zen 3, 4, 5 respectively. \textbf{(right)} Speedup of the online kernel with manual SIMD vectorization compared to the auto vectorized one using the maximally mixed state. }
    \label{fig:runtime-eff-hw-ut}
\end{figure*}

\paragraph{Performance Benchmark}
We compare the runtime of the two algorithms and our baseline \texttt{RandomMeas.jl} in Fig.~\ref{fig:werner_runtimes}. As the online algorithm essentially corresponds to running the batch algorithm $m$ times, it is generally slower. The core idea of the online algorithm, however, is concurrent execution with the data acquisition itself, limiting the significance of a direct performance comparison. 
Both the online and batch algorithms initially show a scaling law of $\mathcal{O}(n2^n)$, as the complexity is $\mathcal{O}(n4^n/p)$, with $p=2^n$. For a single node, we run out of cores to add for $n>7$, thus, observe a transition to $\mathcal{O}(n4^n)$ scaling.

For sufficiently large $N$, the sampled unitaries eventually correspond to the $3^n$ possible measurements. In this regime, our batch implementation scales linearly with the total number of measurements $N_U \cdot N_M$, while our baseline has a fixed cost of $N_U=3^n$. Consequently, for sufficiently large $N$, \texttt{RandomMeas.jl} becomes asymptotically more favorable than repeatedly evaluating the increasingly large KrMM recurrence. However, this asymptotic regime lies beyond realistic measurement counts, which is limited by the cost of data acquisition~\cite{Huang_2020_predicting-many-properties,Elben_2023_randomized-meas-toolbox,Rath_2023_batch-shadows}. Nevertheless, in the regime accessible on today's quantum hardware ($N \approx 10^6 - 10^8$), the parallelism exposed by the KrMM recurrence yields speed-ups ranging from $1.66\times$ ($n=8$, $N=10^8$) to $516\times$ ($n=10$, $N=10^6$).

Fig.~\ref{fig:runtime-eff-hw-ut} (left, brown) shows the strong-scaling efficiency of the distributed online algorithm.
Our row-wise partition ensures that inter-node communication is limited to a final reduction.
Therefore, we observe close-to-ideal strong scaling efficiency; above $0.95$ for all multi-node executions with up to $16$ nodes. At $2$ nodes, we even observe a minor superlinear speed-up, which we attribute to a hardware artifact of the HPC cluster. Note that we only considered power-of-2 node counts P, as it maps perfectly to the row count.
Otherwise, we expect slightly reduced efficiency due to non-ideal load balancing.

We compare the performance of the single-node online and batch algorithms further in Fig.~\ref{fig:runtime-eff-hw-ut} (left), where the pink line compares the runtime of the online algorithm to $m$ times the batched one. For few qubits, the online algorithm is less efficient than the batched one, likely as a result of higher implementation overhead as, despite efficiently fusing the data access, its implementation still involves much bookkeeping. For small $n$, the kernel has a tight execution schedule, and the CPU may struggle amortizing these costs.
For larger systems the online algorithm becomes slightly more efficient.

Having established parallel scalability in a multi-node setting, we turn to single node performance for increasing $n$ and cores, comparing throughput relative to peak performance. Since FMA operations constitute the dominant workload, we take peak FMA throughput as the reference performance
\begin{equation}
    \label{eq:pp-formula}
    \mathrm{freq}(n) \cdot \#\mathrm{cores}(n) \cdot \mathrm{vectorsize}(\mathrm{max}) \cdot \#\mathrm{pipes}(\mathrm{FMA}) \cdot 2,
\end{equation}
where the factor 2 comes from one FMA operation performing two flops.
To obtain a faithful estimate at varying core counts, we measure the CPU frequency during kernel execution using the \texttt{perf} profiling tool. 
Thus, at $n=12$ (full core usage) we arrive at peak performances of $3.96$, $1.77$, $2.18$ TFLOPs for Zen 3, 4, 5 respectively. The performed flops in the dominant FMA section of the row kernel are given as $14 \cdot N \cdot m \cdot n \cdot 4^n$, where $14$ are the flops to compute $\alpha \cdot x + \beta \cdot y$.

Fig.~\ref{fig:runtime-eff-hw-ut} (middle) shows the resulting per-core hardware utilization of the online KrMM kernel for the different microarchitectures. Even though more cores are added for increasing $n$, the per-core hardware utilization does not decrease. Just as in the multi-node setting, all cores work independently, leading to minor parallel overhead. Crucially for the single-node setting, at the considered values of $n$, the data buffers are small enough to fit into either L1 or L2 cache, leading to no heavy bandwidth contention between threads. Due to the compact Kronecker factor representation, loading $\hat{\rho}_t$ constitutes only minor memory traffic per cycle. Both the trivially parallel partition and the high data locality of the row-wise approach thus enable high throughput. 

Initially, at small $n$, utilization is poorer, which continuously rises as $n$ increases. The reasons for this are two-fold: (1) we are limited to smaller vector sizes at $n=2,3$, and (2), at small $n$ the SRM cycle is not repeated many times, leading to implementation overhead dominating total cost. From $n \approx 6$ qubits onward, we plateau at around 57\% for Zen 3, and 75\% for Zen 4 and 5. 
The performance of Zen 5 does, however, decrease to roughly $57\%$ starting at $n>9$ qubits. We attribute this performance drop to a transition from L1 to L2 cache, since this coincides exactly with the row buffers leaking L1. Interestingly, this transition manifests neither in the other architectures nor in the auto-vectorized version on Zen 5.

\begin{figure}[!t]
\centerline{\includegraphics[width=0.325\textwidth]{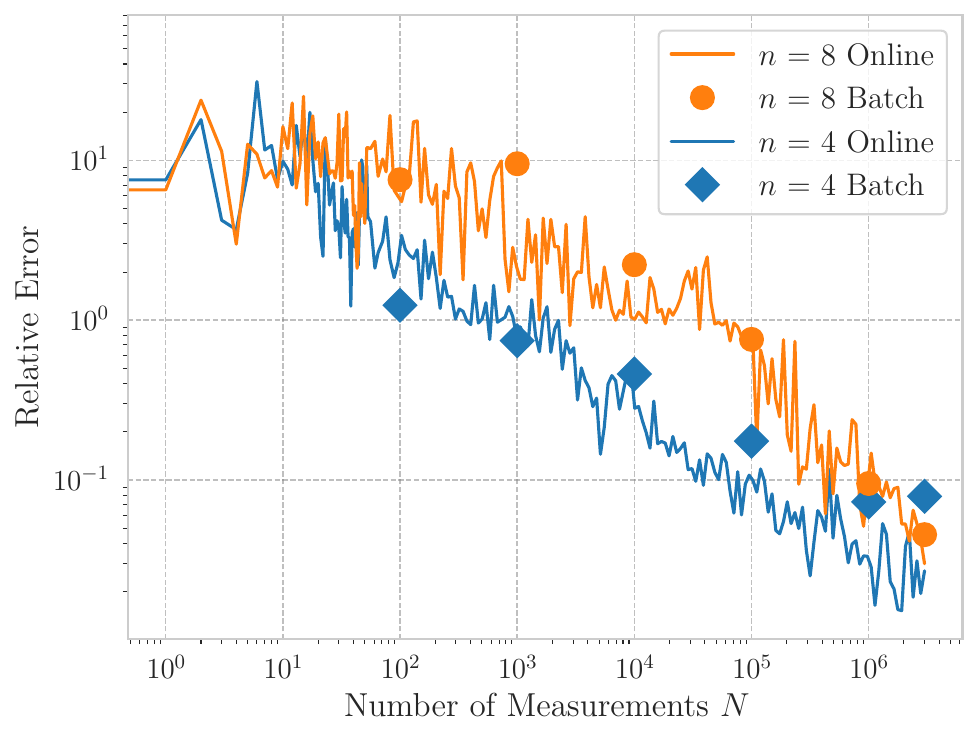}}
\caption{Relative error of estimating the QFI of the random rank-6 state with $n=4$ and $n=8$ qubits. Solid lines depict the online algorithm; dots correspond to the batch algorithm.}
\label{fig:qfi-rel-err}
\end{figure}

Lastly, we show the speed-up of our custom SIMD vectorization to an auto-vectorized implementation in Fig.~\ref{fig:runtime-eff-hw-ut} (right). We implemented the auto-vectorized baseline using the same row kernel execution model of Fig.~\ref{fig:row-kernel}, but used a naive data layout of interleaved real and imaginary floats, and performed the arithmetic according to Eq.~\eqref{eq:vec2}. We can see that our custom vectorization consistently outperforms the auto-vectorized one. It is particularly well suited for AVX-512 architectures (Zen 4, Zen 5), resulting in a rapid increase in speed-up at $n=4$ qubits. The lower speedup of Zen 3, coupled with the lower saturated hardware utilization of $57\%$ implies that the hardware support for SIMD intrinsics shuffling data from arbitrary entries of two vectors to a new vector (as in AVX-512) is crucial for the performance of our SIMD scheme.

\subsection{Quantum Fisher information}
\paragraph{Accuracy and Convergence}
In KST, the rank of the state heavily dictates the effort needed to estimate the QFI. This is formulated in the bound order $N^*$, at which the estimated lower bound and the QFI coincide exactly~\cite{Wang_2026_KST-superiority}. In our experiments, we employ a random rank-6 state, where $N^* = 21$. To estimate $F_Q$ directly, we need to estimate the polynomials up to order $43$, according to Eq.~\eqref{eq:QFI-polys}.

Fig.~\ref{fig:qfi-rel-err} shows the relative error of estimating $F_Q$ of a random rank-6 state with $n=4,8$ qubits for the algorithms. They produce a similar error up to noisy uncertainty, both requiring around $3\cdot 10^6$ measurements for a relative error of $10^{-1}$. While the estimator for $8$ qubits is clearly less accurate than for $4$ qubits, the difference is less pronounced than for the PT moments.
For the QFI, we refer to the true relative error as it is not as straightforward to estimate uncertainty. While we have access to the variance of $q_l^{(k)}$, propagation to $F_Q$ is unclear. The non-polynomial approach lies at the core of KST, however, it also makes uncertainty estimation non-trivial and beyond the scope of this paper. Previously~\cite{Zhang_2026_KST_experimental}, this was addressed by repeating batch shadows 100 times in a bootstrapping scheme over all $N$ measurements.

\begin{figure}[!t]
\centerline{\includegraphics[width=0.325\textwidth]{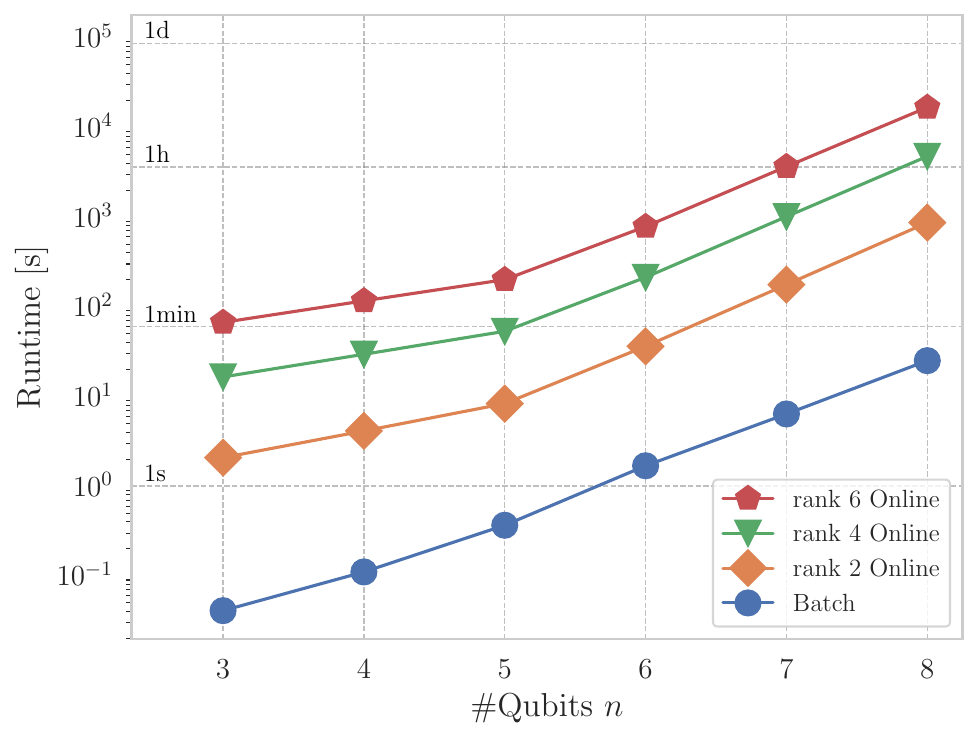}}
\caption{Runtimes of QFI using the random rank 2,4,6 states. Batch uses rank 6. For rank 2 we have $N^* = 3$ and for rank 4 $N^* = 10$. Performed on the Zen 4 configuration.
}
\label{fig:qfi-runtime}
\end{figure}

\paragraph{Performance Benchmark}
Fig.~\ref{fig:qfi-runtime} compares the runtime using $3\cdot 10^6$ measurements for different system sizes. The batched version shows clear advantages, requiring less than $1$ minute even on $8$ qubits. The online algorithm requires $N^* + 1$ independent runs to estimate all polynomials, so it has a clear runtime disadvantage for higher rank states.

\section{Limitations}\label{sec:limitations}
While our approach provides a practical improvement, it does not eliminate the exponential memory and runtime scaling, which is dictated by the exponential sample complexity. 
Nevertheless, the considered problems are highly relevant, especially in the small- to intermediate-scale systems regime for which classical shadows were originally designed. 
Our method, together with the online approach and early stopping from~\cite{marso2026onlineapproachentanglementverification}, extends the range of feasible problem sizes and significantly improves the efficiency of the estimators within the system sizes relevant to current classical shadows applications.

\section{Conclusion and Future Work}\label{sec:concl-future-work}
In this paper, we identified a recurring pattern in classical shadow estimators, allowing for significant practical speed-ups on practically relevant but computationally difficult problems.
We formulated the Kronecker-matrix-matrix recurrence, resulting in independent row-wise partitions that allow parallelizing the execution.
In combination with a custom SIMD vectorization, it enables optimizing the throughput and parallel scaling.

Further, we have shown that the structure naturally occurs in estimating polynomial functions in classical shadows applications, with the examples of mixed state entanglement verification and the estimation of the Quantum Fisher Information.
We defined online estimators that continuously monitor convergence, allowing for advanced hybrid workflows. In this approach, no dense matrix ever needs to be stored concurrently, leading to reduced memory requirements.

More broadly, our framework can be applied to batch shadows, the state-of-the-art estimator for polynomial functions in an offline setting.
This enables parallelism beyond the conventional concurrent execution of batch computation, translating into significant speed-ups over the reference implementation. 

With the establishment of online algorithms for classical shadows it would be an interesting future avenue to develop a pipeline for an actual online setting. We have not considered this in this work due to (1) limited availability of quantum hardware and (2) limited support to-date from hardware providers for online hybrid algorithms. Further, in light of noisy quantum measurements, it would be interesting to develop mitigation strategies, e.g., through robust shadows~\cite{Chen_2021_robust-shadows}.

\section*{Acknowledgments} 
For this work, we have used AI to improve reading flow and clarity, however, not for text generation. Further, we used it to aid with visualization. We have reviewed all AI-assisted work by manually verifying correctness and take full responsibility for the final content of this work. 
\\

The computational results have been achieved using the Austrian Scientific Computing (ASC) infrastructure.
SH is a recipient of the Andreas Dieberger-Peter Skalicky scholarship awarded by TU Wien and acknowledges financial support by the netidee Austria funding campaign (Grant no. 7728). This work has been partially funded through the Themis project (Trustworthy and Sustainable Code Offloading), Austrian Science Fund (FWF) (Grant no. 10.55776/PAT1668223), the Standalone Project Transprecise Edge Computing (Triton), Austrian Science Fund (FWF) (Grant no. P 36870-N), BMIMI, BMWET, the State of Upper Austria, and the State of Tyrol within the COMET module Quantum Algorithm Engineering (FFG grant no. 923923) managed by Austrian Research Promotion Agency FFG, and by the Advanced learning paradigms for Photonic quantum processors (ALPAQA) project, funded by the Vienna Science and Technology Fund (WWTF) (Grant no. 10.47379/ICT25062).

\bibliographystyle{IEEEtran}
\bibliography{bib/references}

\end{document}